\documentclass[11pt,twoside]{article}

\usepackage{asp2006}
\usepackage{graphicx}

\begin{document}
\title{Reflections of Cluster Assembly in the Stellar Populations and
  Dynamics of Member Galaxies}   %%% Fill in title
\author{Sean M. Moran$^1$, Richard S. Ellis$^1$, Tommaso Treu$^2$, Graham P. Smith$^3$}   %%% Fill in author names
\affil{$^1$Caltech, Dept. of
  Astronomy, MC 105-24, Pasadena, CA 91125,
  $^2$Dept. of Physics, University of California, Santa Barbara,
  CA 93106,
  $^3$University of Birmingham,
  Edgbaston, Birmingham, B15 2TT, UK}    %%% Fill in author affiliations

\begin{abstract} %%% Abstract to run on from here.
We combine optical (HST) and UV (GALEX) imaging of two
intermediate redshift galaxy clusters with spectroscopy of member
galaxies, to study the relation between the
formation history of cluster galaxies and the assembly history of
the cluster structure itself. 
We identify key differences in the large-scale structure and
intracluster medium properties of each 
cluster.
In order to assess the importance of cluster
substructure and the ICM in the evolution of cluster galaxies, we examine
several key indicators of the recent star-formation and assembly 
history of cluster galaxies. We find that galaxies 
in cluster MS~0451 ($z=0.54$) exhibit a markedly lower incidence of 
recent star formation 
activity than galaxies in cluster Cl~0024 ($z=0.39$), 
likely the result of starvation by
the ICM. In addition, Cl~0024 members show evidence for kinematic
disturbances that can be linked to the assembly of substructure. 
\end{abstract}
%%% MAIN BODY OF TEXT GOES HERE. CONSULT "INSTRUCTIONS FOR AUTHORS USING
%%% LATEX2E MARKUP", SECTIONS 2.3-2.6 FOR HELP WITH EQUATIONS, FIGURES,
%%% AND TABLES.
%\section{}   %%% Top level section head (remove "%" symbol)
%\subsection{}   %%% Second level section head (remove "%" symbol)
%\subsubsection{}   %%% Lowest level section head (remove "%" symbol)
%\section*{}    %%% Unnumbered top level section head (remove "%" symbol)
%\subsection*{}   %%% Unnumbered second level section head (remove "%" symbol)
\section{INTRODUCTION}
In general, it is well-known that environmental processes play a 
significant role in shaping the evolution of galaxies as they assemble
onto clusters. With the aid of Hubble 
Space Telescope ({\it HST}) imaging and deep optical spectroscopy, 
recent studies have quantified this evolution in galaxy properties, 
painting a picture where the fraction of early-type 
(elliptical and S0) galaxies and the fraction of passive
non-star-forming galaxies both grow with time, and at a rate that
seems to depend sensitively on the local density of galaxies
\citep{smith05b,postman05,poggianti06}.

Yet there are a wide variety of physical processes that may be
responsible for these evolutionary trends--including galaxy mergers, 
harassment, gas stripping by the ICM, or tidal processes 
\citep[e.g.][]{moore99,fujita98, bekki02}. 
Observationally, it has so far been impossible
to fully separate the effects of the various physical processes, 
in large part due to the overlapping regimes 
of influence for each of the proposed mechanisms \citep[see][]{tt03}.
Further complicating the picture, the large scale
assembly states of clusters show considerable variety
\citep{smith05a}, such that the dominant forces acting on galaxies are 
likely to vary from cluster to cluster, or over the course of an 
individual cluster's assembly history.  
But gaining an understanding of the complex interplay between a
variable ICM, the properties of assembling galaxies, and the overall 
cluster dynamical state is crucial if we are to have a complete picture of the
growth and evolution of galaxies in a hierarchical universe.

In this contribution, we combine optical (HST) and UV
(GALEX) imaging of two $z\sim0.5$ galaxy clusters with 
ground-based (Keck) spectroscopy of member galaxies, in an attempt to
better characterize the relation between the formation history of 
cluster galaxies and the assembly history of the cluster structure itself. 
\section{OBSERVATIONS}
We make use of {\it HST} imaging of Cl~0024
and MS~0451 from the comprehensive wide-field survey described in
\citet{tt03} and Smith et al. (2007, in
preparation). In Cl~0024, HST coverage consists of a sparsely-sampled
mosaic of 39 WFPC2 images taken in the {\it F814W} filter ($\sim I$ band),
providing coverage to a projected radius $> 5$ Mpc. MS~0451
observations were taken with the ACS, also in {\it F814W}, and 
provide contiguous coverage
within a 10Mpc$\times$10Mpc box centered on the cluster. 
For both clusters, reliable morphological classification is possible to 
rest frame absolute magnitude $M_V=-19.5$. 
All galaxies brighter than this limit are classified visually
following \citet{tt03}.

Cl~0024 and MS~0451 were respectively observed for 15ksec and 80ksec
with {\it GALEX} \citep{martin05} in 2004 October 
(GO-22; Cycle 1; PI Treu), reaching comparable depths in rest frame
{\it FUV} (observed {\it NUV}). Galaxy fluxes were measured within $6\arcsec$
circular apertures, centered on the optical position, and comparable
to the measured {\it NUV} FWHM ($5\farcs5$).

Observations with the DEIMOS spectrograph on Keck II from October 2001
to October 2005 secured spectra for over $500$ members of both
Cl~0024 ($0.373 < z < 0.402$) and MS~0451 ($0.520 < z < 0.560$). 
Details are provided in
\citet{moran05} and Moran et al. (2007, in preparation). Briefly, we observe with
$1\arcsec \times 8\arcsec$ slits, with a typical velocity resolution of 
50~km~s$^{-1}$, covering rest frame wavelengths
from $\sim3500\mbox{\AA}$ to $\sim6700\mbox{\AA}$. 
Exposure times totaled 2.5hrs in Cl~0024 and 4hrs in
MS~0451.
DEIMOS data were reduced using the DEEP2 data reduction
pipeline \citep{davis03}.

\section{RESULTS}

\subsection{Global Cluster Properties}

\begin{table}[b]

  \smallskip
  \begin{center}
  \caption{Basic Properties of the Clusters}
  {\small
\begin{tabular}{llccc}
  \tableline 
  \noalign{\smallskip}
 Name  & $R_{VIR}$ & $M_{200}$ & z & $L_X$ \\
  & (Mpc) & ($M_\odot$) & & ($L_\odot$) \\
  \noalign{\smallskip}
\tableline
  \noalign{\smallskip}
Cl~0024 & $1.7^{(1)}$ & $8.7\times 10^{14} \ ^{(2)} $ & 0.395 & $7.6 \times
10^{10} \ ^{(3)}$ \\
MS~0451 & 2.6  & $1.4\times 10^{15} \  ^{(4)}$ & 0.540 & $3.8 \times
10^{11} \  ^{(4)}$\\
  \noalign{\smallskip}
\tableline
\end{tabular}
}

{\tiny \centering $^{(1)}$ Treu et al. (2003), $^{(2)}$ Kneib et al. (2003), $^{(3)}$
  Zhang et al. (2005), $^{(4)}$ Donahue et al. (2003).}
\end{center}
\end{table}

Similar in their total mass, the two
clusters were chosen for study primarily due to their
complementary X-ray properties (See Table~1). 
While MS~0451 is one of the most X-ray
luminous clusters known \citep{donahue03}, Cl~0024 is somewhat
under-luminous \citep{zhang05}. This implies a large difference in the density and
radial extent of the intracluster medium (ICM) between the two
clusters, such that ICM-related physical processes are expected to be
significantly more important in the evolution of MS~0451 galaxies.
In addition, marked differences in the spatial and redshift distributions of
galaxies between the two clusters indicate that the clusters have quite
dissimilar recent assembly histories. While the distribution of
galaxies in MS~0451 is largely smooth and seemingly well-Virialized,
Cl~0024 shows obvious signatures of infalling groups \citep{tt03} and,
indeed, a recent cluster-subcluster merger \citep{czoske01}. In the
following, we examine how such differences in the overall cluster
environment lead to striking differences in the galaxies themselves.

\subsection{Passive Spirals}

Passive spirals belong to a class of galaxies that show spiral
morphology in {\it HST} images, but reveal weak or no [OII]
emission in their spectra, suggesting a lack of current star
formation.  Through GALEX UV imaging, we find that passive 
spirals in Cl~0024 exhibit UV emission nearly as strong as regular 
star-forming spirals, implying the presence of young
stars \citep{moran06}. Their unusual 
combination of UV emission with weak H$\delta$ strength 
supports a picture where passive spirals have experienced a rapid 
decline and eventual cessation of star formation over the last $\sim1$~Gyr.
The timescale of this decline implicates ``starvation'' by the ICM as
the likely cause \citep{bekki02}--a process where diffuse gas is
stripped from a galaxy's halo, thereby halting any further accretion of
cold gas onto the galactic disk.

In MS~0451, over one third of all spiral galaxies fall into the
`passive' category, compared to $\sim24\%$ in Cl~0024. Of these, $1/3$
are detected in the UV (c.f. $2/3$ in Cl~0024), 
suggesting that the typical passive spiral in
MS~0451 has fewer young stars than those in Cl~0024. Both effects are
likely due to the denser, more extended ICM in MS~0451, which should
boost the efficiency of starvation, and increase rate of passive 
spiral creation compared to that of Cl~0024. In fact, the spatial 
distribution of passive spirals across both clusters
strongly supports this scenario (Figure~1). Passive spirals
in MS~0451 are spread across a wide area, as expected
in the presence of a dense ICM, while Cl~0024
passive spirals are largely concentrated near the cluster core, where
its ICM is densest.

\subsection{Kinematic Disturbances}

 By constructing the Fundamental Plane (FP) of Cl~0024, we observe that
 elliptical and S0 galaxies (E+S0s) exhibit a 
 high scatter in their FP residuals, equivalent
 to a spread of $40\%$ in mass to light ratio ($M/L_V$) \citep{moran05}.
Upon closer inspection, this high scatter appears to occur only among
 galaxies within 1~Mpc of the cluster core. Outside of this radius,
 cluster E+S0s follow a tight FP. In contrast, ellipticals in MS~0451 do not
 exhibit such a high scatter, and no radial dependence is seen. It is
 therefore likely that the recent merger with a subcluster
 \citep{czoske01} has disturbed galaxies in the core of Cl~0024, such
 that they show significant deviations from the FP. 

Similarly, in a recent study of spiral galaxies across both clusters 
\citep{moran07a}, a large scatter about the
 Tully-Fisher relation is measured, compared to the field at
 the same redshift. In \citet{moran07a}, we argue that the large
 scatter is likely driven by galaxy--galaxy interactions during
 cluster infall.

\section{DISCUSSION}

\begin{figure}[t]
\includegraphics[width=0.5\columnwidth]{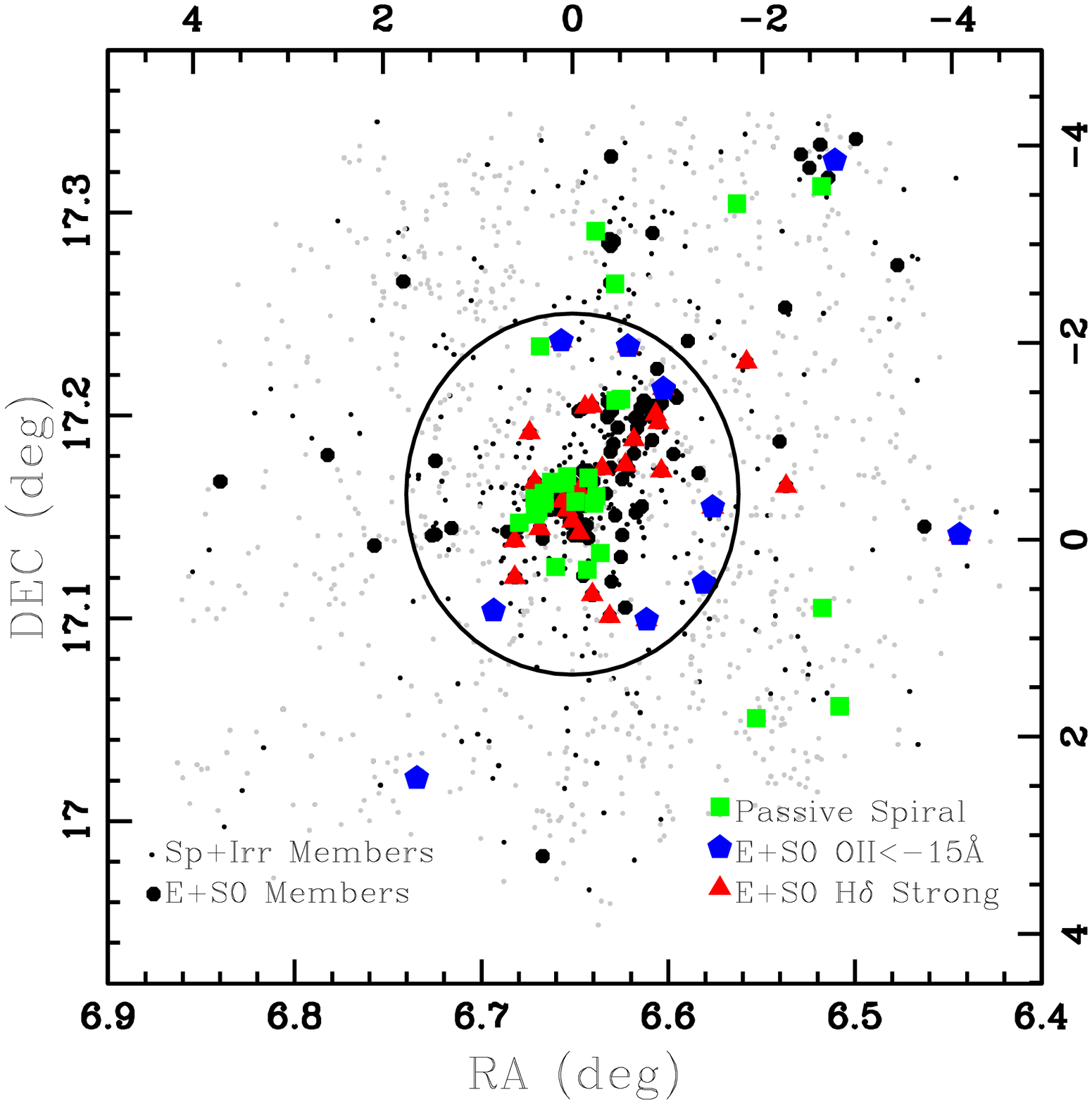}
\includegraphics[width=0.5\columnwidth]{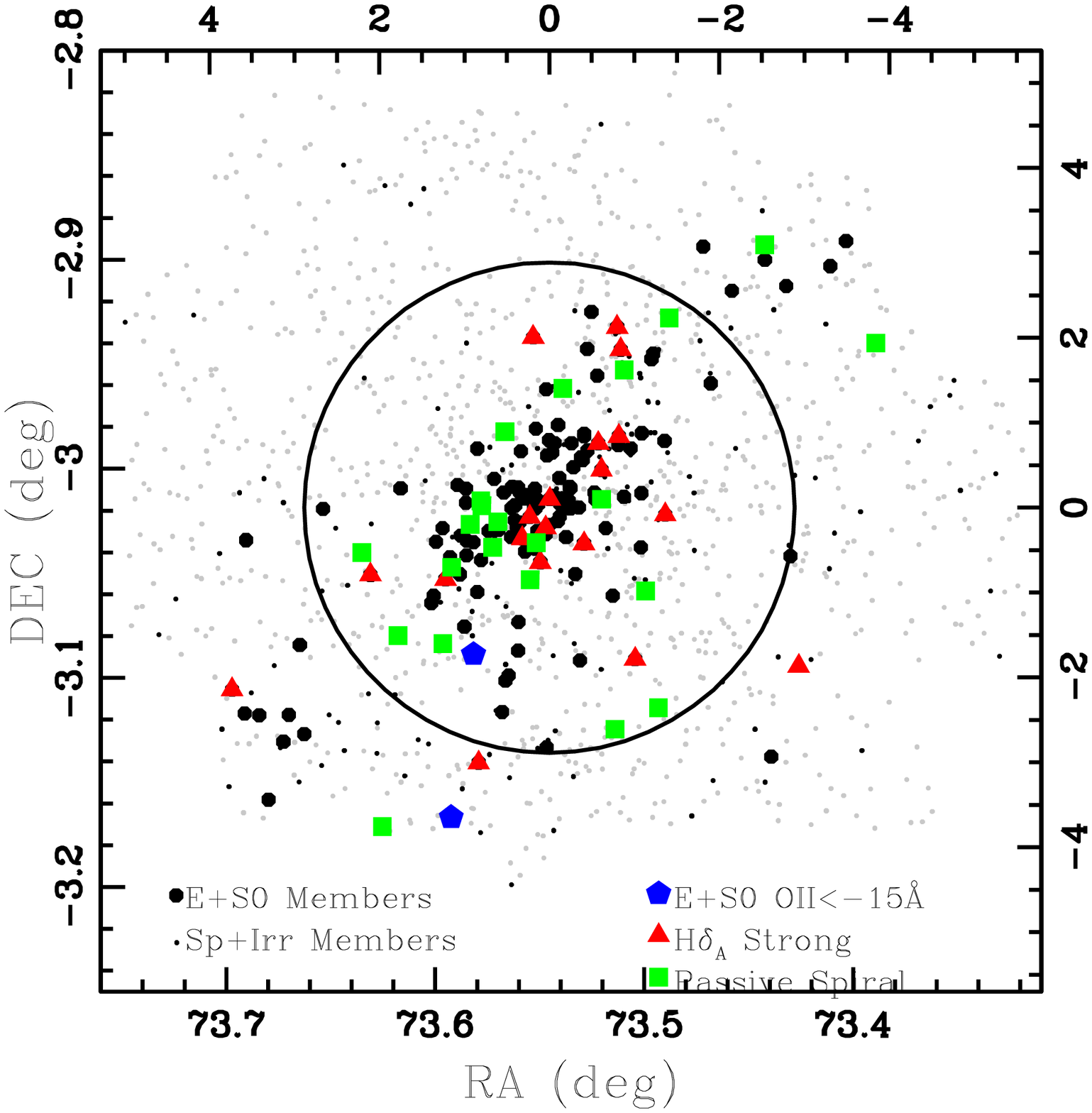}

\caption{Distribution of galaxies in the field
  of Cl~0024 (left) and MS~0451 (right). 
  Galaxies showing spectroscopic signs of recent 
  evolution are marked. 
  The distributions of passive spirals are 
  quite different between the two clusters. 
}
\end{figure}

Taken together, our results paint a picture where the fates of
infalling cluster galaxies depend heavily on the pre-existing assembly
state and ICM density of the cluster. Cluster--subcluster mergers, as
seen in Cl~0024, seem to induce departures from the FP, and cluster
spirals show signs of similar disturbances. 
Importantly, passive spirals appear to be surprisingly abundant, 
suggesting that starvation by the
ICM causes a steady decline in star formation rate, driving the
eventual buildup of the cluster red sequence. 

%\acknowledgements %%% Text of acknowledgements runs on after this command.

%%% THE BIBLIOGRAPHY
%%%
%%% CONSULT SECTION 3 OF "INSTRUCTIONS FOR AUTHORS" FOR HOW TO USE NATBIB.
%%% AUTHORS ARE ENCOURAGED TO USE EITHER THE "THEBIBLIOGRAPY" ENVIRONMENT
%%% BY UNCOMMENTING (DELETING THE "%" SYMBOL) THE COMMANDS BELOW, OR BY
%%% USING THE BIBTEX ENVIRONMENT. TO FIND OUT WHICH IS APPLICABLE TO YOUR
%%% CONTRIBUTION, CONSULT THE VOLUME EDITORS FOR YOUR PROCEEDINGS.
%%%


\begin{thebibliography}{}
\bibitem[Bekki et al.(2002)]{bekki02} Bekki K., Couch, W.J. \& Shioya,
  Y. 2002, ApJ, 577, 651
\bibitem[Czoske et al.(2001)]{czoske01}Czoske, O. et al. 2001, A\& A, 372, 391
\bibitem[Davis et al.(2003)]{davis03} Davis, M. et al. 2003, SPIE, 4834, 161
\bibitem[Donahue et al.(2003)]{donahue03} Donahue, M. et al. 2003, ApJ, 598, 190
\bibitem[Fujita(1998)]{fujita98} Fujita, Y. 1998, ApJ, 509, 587
\bibitem[Kneib et al.(2003)]{kneib03} Kneib, J-P. et al. 2003, \apj, 598, 804
\bibitem[Martin et al.(2005)]{martin05} Martin, D. C. et al. 2005, ApJ, 619, L1
\bibitem[Moore et al.(1999)]{moore99} Moore, B. et al. 1999, MNRAS, 304, 465
\bibitem[Moran et al.(2007a)]{moran07a} Moran, S. M. et al. 2007, ApJ
  in press
\bibitem[Moran et al.(2006)]{moran06} Moran, S. M. et al. 2006, ApJ, 641, L97
\bibitem[Moran et al.(2005)]{moran05} Moran, S. M. et al. 2005, ApJ, 634, 977
\bibitem[Poggianti et al.(2006)]{poggianti06} Poggianti, B. et al. 2006, ApJ, 642, 188
\bibitem[Postman et al.(2005)]{postman05} Postman, M. et al. 2005, ApJ, 623, 721
\bibitem[Smith et al.(2005a)]{smith05a} Smith, G. P. et al. 2005a, MNRAS, 359, 417
\bibitem[Smith et al.(2005b)]{smith05b} Smith, G. P. et al. 2005b, ApJ, 620, 78
\bibitem[Treu et al.(2003)]{tt03} Treu, T. et al. 2003, ApJ, 591, 53
\bibitem[Zhang et al.(2005)]{zhang05} Zhang, Y. -Y. et al. 2005, A\& A, 429, 85
\end{thebibliography}
\end{document}